\documentclass[%
 reprint,
 amsmath,amssymb,
 aps,prd
floatfix,
showkeys
]{revtex4-2}
\usepackage{lmodern}
\usepackage{xcolor}
\usepackage{ulem}
\usepackage{graphicx}
\usepackage{dcolumn}
\usepackage{bm}
\usepackage{hyperref}

\usepackage{float}
\usepackage{physics}
\usepackage{tikz}
\usepackage{tikz-feynman}
\usepackage[caption = false]{subfig}
\usepackage{enumerate}
\usepackage{enumitem}
\usepackage{siunitx}

\newcommand{\fii}{\varphi}								
\renewcommand{\a}{\alpha}								
\renewcommand{\b}{\beta}								
\newcommand{\g}{\gamma}									
\renewcommand{\d}{\delta}								
\newcommand{\e}{\varepsilon}

\renewcommand{\epsilon}{\varepsilon}					
\renewcommand{\t}{\theta}

\newcommand{\ds}{\displaystyle}

\graphicspath{{plots/}}

\begin{document}


\title{Electric corrections to $\pi$-$\pi$ scattering lenghts in the linear sigma model}

\author{R. Cádiz}
\affiliation{Facultad de F\'isica, Pontificia Universidad Cat\'olica de Chile, Vicu\~{n}a Mackenna 4860, Santiago, Chile.}

\author{M. Loewe}%
\affiliation{Facultad de Ingeniería, Arquitectura y Diseño, Universidad San Sebastián, Santiago, Chile.}
\affiliation{Centre for Theoretical and Mathematical Physics, and Department of Physics, University of Cape Town, Rondebosch 7700, South Africa}
\author{R. Zamora}
\affiliation{Instituto de Ciencias B\'asicas, Universidad Diego Portales, Casilla 298-V, Santiago, Chile.\\
Facultad de Medicina Veterinaria, Universidad San Sebasti\'an, Santiago, Chile.}


\begin{abstract}
In this article we analyze the role of an external electric field, in the weak field approximation, on $\pi$-$\pi$ scattering lengths. The discussion is presented in the frame of the linear sigma model.  To achieve this, we take into account all one-loop corrections in the $s$, $t$, and $u$ channels associated with the insertion of a Schwinger propagator for charged pions, focusing on the region characterized by small values of the electric field. Furthermore, one of the novelties  of our work is the explicit calculation of box diagrams, which were previously overlooked in discussions regarding magnetic corrections. It turns out that the electric field corrections have an opposite effect with respect to magnetic corrections calculated previously in the literature.
\end{abstract}

\maketitle


\section{\label{sec:level1}Introduction}

During the last years, several research proposals have concentrated their attention on the response of matter under extreme conditions due to external agents. A clear example of this corresponds to relativistic heavy ion collisions, like Au-Au, where extreme high temperatures are produced, allowing for different phase transitions like deconfinement or chiral symmetry restoration. Density effects might be also taken into account, producing an interesting and rich phase diagram in the temperature-density plane. The latter scenario can be found in compact objects like neutron stars. Also, a quite interesting role is played by the extremely high magnetic field produced during the very first stages of the collision between two heavy nuclei like Au-Au. In the case of a relativistic collision between a heavy and a light nuclei, for example Au-Cu, also an electric field appears, due to the imbalance in the number of protons associated to each nucleus. Both fields, the electric and the magnetic one, are produced in a relative perpendicular configuration. The magnetic field points essentially perpendicular to the collision plane, whereas the electric field points along the collision plane.  In the existing literature, there are several works related to the study of different physical parameters in the presence of a magnetic and/or electric field \cite{Gusynin:1995nb,Miransky:2002rp,Shovkovy:2012zn,Bali:2012zg,Bali:2014kia,Kogut:2002zg,Bazavov:2017dus,Johnson:2008vna,Frasca:2011zn,Andersen:2021lnk,Bandyopadhyay:2020zte,G1,Zayakin:2008cy,Filev:2009xp,Ballon-Bayona:2020xtf,G2,G3,G4,G5,fernandez}.

It is important to elucidate the relative effects associated to these external agents on physical, in principle measurable, quantities. For example we know that temperature and magnetic field conspire again each other in several scenarios. Here we want to consider the effect of an external weak electric field on $\pi $-$\pi $ scattering lengths, being these pions produced during the collision.
For this purpose we will work in the frame of the linear sigma model, computing all relevant corrections to the scattering lengths. Fermion contributions are neglected in the analysis since they are much more massive than the other particles involved in the model: pions and the scalar sigma field. It is important to avoid the strong field case since Schwinger instabilities associated to pair productions of pions could appear. This scenario goes beyond the present discussion.

 Our discussion also considered the effects of box diagrams, previously not taken into account \cite{scattering1,scattering2,scattering3} due the the relative high mass of the sigma field.  Although the influence of those diagrams, as expected, is in fact small, they could play a relevant role when summing up, for example, ladder diagrams with  boxes, looking for reggeized amplitudes dependent on external agents.

This analysis of $\pi$-$\pi$ scattering lengths corrections has been carried out previously, using the same model, for the magnetic case \cite{scattering1,scattering2,scattering3}. As we will see, the electric field corrections turn out to be opposite respect to the equivalent magnetic corrections. This is an indication that it will be not an easy task to isolate, in a clear way, the influence a certain specific external agent on physical observables.

This article is organized as follows: In section \ref{sec2} we present the linear sigma model concentrating on the general structure of the $\pi $-$\pi $ scattering amplitudes an their projection into different isospin channels. In section \ref{sec3} we present the propagator for a boson immersed in a constant external electric field showing also its weak field limit. We then  go into the different relevant diagrams and the techniques used for their evaluation. Details are given in the appendices. Finally, we present our conclusions.

\section{\label{sec2}Linear sigma model and $\pi$-$\pi$ Scattering}

 Gell-Mann and Lévy \cite{Gell-Mann:1960mvl} proposed the linear sigma model (LSM) as an effective framework to elucidate chiral symmetry breaking through both explicit and spontaneous mechanisms. In the chiral broken phase the model is expressed as
\begin{eqnarray}
&&\mathcal{L}=\bar{\psi}\left[i\gamma^{\mu}\partial_{\mu}-m_{\psi}-g(\sigma+i\vec{\pi}\cdot\vec{\tau}\gamma_{5})\right]\psi \nonumber \\
&&+\frac{1}{2}\left[(\partial\vec{\pi})^2+m_{\pi}^2\vec{\pi}^2\right]+\frac{1}{2}\left[(\partial\sigma)^2+m_{\sigma}^2 \sigma^2\right]\nonumber\\
&&-\lambda^2v\sigma(\sigma^2+\vec{\pi}^2)-\frac{\lambda^2}{4}(\sigma^2+\vec{\pi}^2)^2+(\varepsilon c-vm_{\pi}^2)\sigma.
\end{eqnarray}
Pions are described by an isospin triplet, $\vec{\pi}=(\pi_1,\pi_2,\pi_3)$, $c\sigma$ is the term that breaks explicitly the $SU(2) \times SU(2)$ chiral symmetry, being $\sigma$ a scalar field, and $\varepsilon$ is a small dimensionless parameter. The model also incorporates a doublet of Fermi fields, associated in the original version to nucleon states, which in our context  will be ignored since they are too heavy as compared to the scalar sigma meson mass and to the relevant energy scale. It is intriguing to note that the masses of all fields in the model are determined by $v$. Indeed, the following relations can be proved to be valid:
$m_{\psi}=gv$, $m_{\pi}^2=\mu^2+\lambda^2v^2$ and
$m_{\sigma}^2=\mu^2+3\lambda^2v^2$. 
Perturbation theory at the tree level allows us to identify the pion decay constants as $f_{\pi}=v$. 

The LSM turns out to be a wonderful scenario for exploring effects of external agents like temperature, magnetic field, electric field, and vorticity. These effects have been studied in a series of articles by various authors, ~\cite{wagner,kovacs2,kovacs1,kovacs3}. In the present work we will explore, in the frame of the LSM model, how an external electric field, generated in collisions between  heavy and a light nuclei, as for example  Au-Cu collisions, will affect the $\pi$-$\pi$ scattering lengths. We will compare our results with previous analysis where instead a magnetic field was considered.

The most general decomposition for the scattering amplitude for particles with definite isospin quantum numbers is given by~\cite{Collins, gasser}

\begin{eqnarray}
T_{\alpha\beta;\delta\gamma}&=&A(s,t,u)\d_{\a\b}\d_{\g\e}+A(t,s,u)\d_{\a\e}\d_{\b\g}\nonumber\\
&&+A(u,t,s)\d_{\a\g}\d_{\b\e},
\label{proyectores}
\end{eqnarray}
\noindent where $\alpha$, $\beta$, $\gamma$, $\delta$ represent isospin components.

Through the utilization of suitable projection operators
\begin{align}
P_0&=\dfrac{1}{3}\d_{\a\b}\d_{\g\e}\label{ProjOp0},\\
P_1&=-\dfrac{1}{2}\qty(\d_{\a\g}\d_{\b\e}-\d_{\a\e}\d_{\b\g})\label{ProjOp1},\\
P_2&=\dfrac{1}{2}\qty(\d_{\a\g}\d_{\b\e}+\d_{\a\e}\d_{\b\g}-\dfrac{2}{3}\d_{\a\b}\d_{\g\e})\label{ProjOp2},
\end{align}
it is possible to find the following isospin dependent scattering amplitudes

\begin{align}
T^{0}&=3A(s,t,u)+A(t,s,u)+A(u,t,s),\label{eq3}\\
T^{1}&=A(t,s,u)-A(u,t,s),\label{eq4}\\
T^{2}&=A(t,s,u)+A(u,t,s),
\label{eq5}
\end{align}

\noindent where $T^I$ denotes a scattering amplitude in a given isospin channel $I = \{0,1,2\}$.\\

As is commonly understood~\cite{Collins}, below the inelastic threshold any scattering amplitude can be expanded in terms of partial amplitudes parameterized by phase shifts for each angular 
momentum channel $\ell$. Hence, in the low-energy region the isospin dependent scattering amplitude can be expanded in partial wave components $T_\ell^I$. The real part of such an amplitude
\begin{equation}
\Re\left(T_{\ell}^{I}\right)=\left(\frac{p^{2}}{m_{\pi}^{2}}\right)^{\ell}\left(a_{\ell}^{I}+\frac{p^2}{m_{\pi}^{2}}b_{\ell}^{I}+\ldots\right),
\end{equation}
is normally expressed in terms of the
scattering lengths $a_{\ell}^{I}$, and the scattering slopes $b_{\ell}^{I}$, respectively.
The scattering lengths satisfy the hierarchy $|a_{0}^{I}|>|a_{1}^{I}|>|a_{2}^{I}|...$.
Specifically, in order to obtain the scattering lengths $a_0^I$, it is
sufficient to calculate the scattering amplitude $T^I$ in the static
limit, i.e., when $s \to 4m_\pi^2$, $t\to 0$ and $u\to 0$,
\begin{equation}
a_{0}^{I}=\frac{1}{32\pi}T^{I}\left(s \to 4m_{\pi}^2,t\to 0, u\to0\right).\label{eq:a0I}
\end{equation} 
 The first measurement of $\pi $-$\pi $ scattering lengths was carried on by Rosellet et al.
~\cite{Rosselet}. More recently, these parameters have been measured  using pionium atoms in the DIRAC experiment \cite{adeva}, as well as through the decay of heavy quarkonium states into $\pi $-$\pi $ final states, where the so called cusp-effect was found~\cite{Liu:2012dv}.

\section{\label{sec3}Scattering lengths at finite electric field}

In previous works, we have computed the magnetic and thermal dependence of the $\pi$-$\pi$ scattering lengths within the framework of the linear sigma model \cite{scattering1,scattering2,scattering3}. In this instance, our aim to employ the same model exploring the electric field  dependence of the $\pi$-$\pi$ scattering lengths. For this purpose, we will use the bosonic scalar propagator in the presence of an electric field \cite{propagadorelectrico1,propagadorelectrico2}, given by

\begin{equation}D(p) =  \int _{0} ^{\infty}{ds}\frac{e^ {-s\left(\frac{\tanh (qiEs)}{qiEs} p_{\parallel}^2 +p_{\perp}^2+ m^2 \right)}}{\cosh (qiEs)}, \label{propE}
\end{equation} 

\noindent
where $q$ is the electric charge, $p_{\parallel}$ and $p_{\perp}$
refer to $(p_{4},0, 0, p_{3})$ and $(0, p_{1},p_{2},0)$, respectively. For simplicity, the electric field points along the z-axis. Note that in the euclidean version $p^{2} = p_{\parallel}^2 + p_{\perp}^2$=$p_4^2+p_3^2+p_1^2+p_2^2$. 
We are interested in the weak electric field region, since for a strong electric field the Schwinger effect might appear, i.e., the generation of electron-positron pairs. Therefore we will proceed to expand the previous expression up to order $\mathcal{O}(E^2)$, to obtain 
\begin{eqnarray}
&&D(p)\approx \frac{1}{p^2+m^2} \nonumber \\
&&-(qE)^2\left(-\frac{1}{(p^2+m^2)^3}+\frac{2  p_{\parallel}^2}{(p^2+m^2)^4}\right).
\label{PiPropagator}
\end{eqnarray}
However, using the relation $p^2=p_\parallel^2+p_\perp^2$, Eq.~\eqref{PiPropagator} can be written as 
\begin{align}
D(p)&\approx\dfrac{1}{p^2+m^2}+\dfrac{q^2E^2\qty[2\qty(p_\perp^2+m^2)-\qty(p^2+m^2)]}{\qty(p^2+m^2)^4}.
\label{PiPropagatorModificado}
\end{align}
The above expression, being more symmetric, is useful to carry on  the integrals that will appear when computing all relevant loop corrections. Also, to distinguish between the free and charged propagators, we will define the free scalar propagator as
\begin{equation}
S(p)=\dfrac{1}{p^2+m^2}.
\label{libre}
\end{equation}



\subsection{\label{sec3a}Loop Integrals Classification}
\begin{figure}[hbtp]
\centering
\subfloat[]{\includegraphics[scale=0.35]{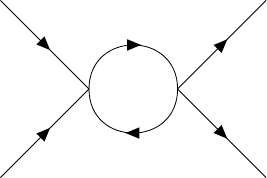}} \hspace*{6mm}
\subfloat[]{\includegraphics[scale=0.35]{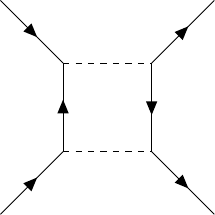}} \hspace*{6mm}
\subfloat[]{\includegraphics[scale=0.35]{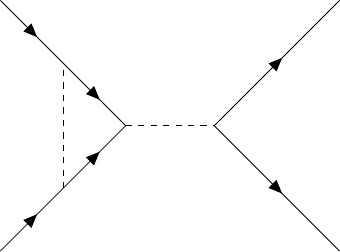}} \\
\subfloat[]{\includegraphics[scale=0.35]{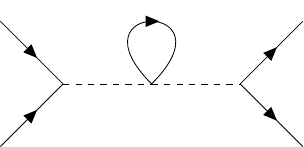}} \hspace*{6mm}
\subfloat[]{\includegraphics[scale=0.35]{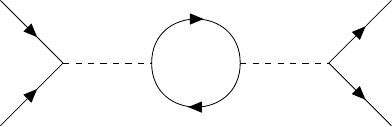}} \hspace*{6mm}
\subfloat[]{\includegraphics[scale=0.35]{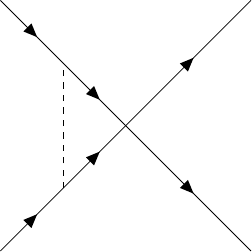}} \\
\subfloat[]{\includegraphics[scale=0.35]{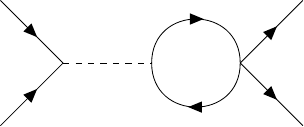}}\hspace*{6mm}
\subfloat[]{\includegraphics[scale=0.35]{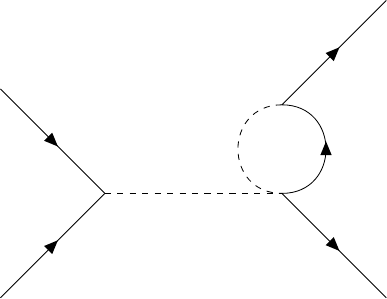}} \hspace*{6mm}
\subfloat[]{\includegraphics[scale=0.35]{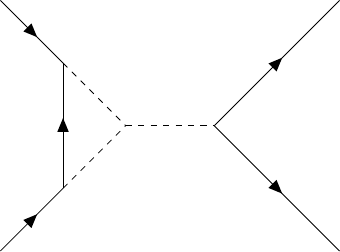}} \\
\subfloat[]{\includegraphics[scale=0.35]{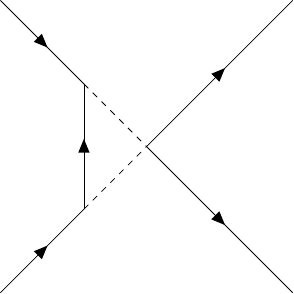}}
\caption{$s$ channel diagrams.}
\label{fig:schanneldiagrams}
\end{figure}

\begin{figure}[hbtp]
\centering
\subfloat[]{\includegraphics[scale=0.35]{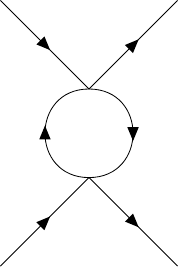}} \hspace*{8mm}
\subfloat[]{\includegraphics[scale=0.35]{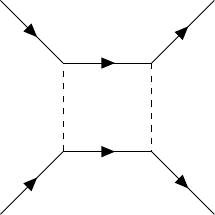}} \hspace*{8mm}
\subfloat[]{\includegraphics[scale=0.35]{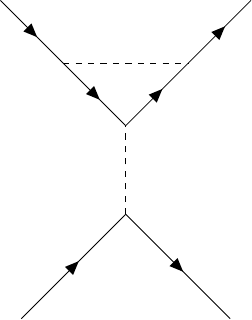}} \hspace*{8mm}
\subfloat[]{\includegraphics[scale=0.35]{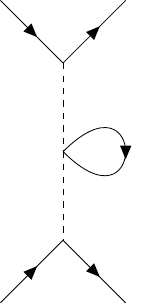}} \\
\subfloat[]{\includegraphics[scale=0.35]{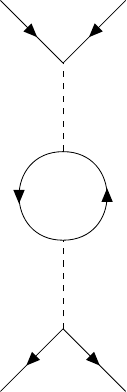}} \hspace*{8mm}
\subfloat[]{\includegraphics[scale=0.35]{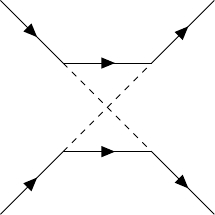}} \hspace*{8mm}
\subfloat[]{\includegraphics[scale=0.35]{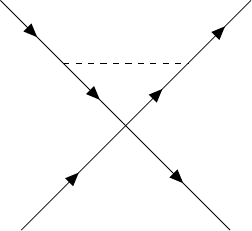}}\hspace*{8mm}
\subfloat[]{\includegraphics[scale=0.35]{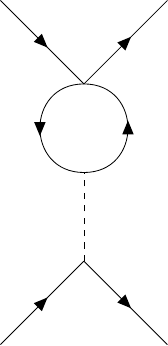}} \\
\subfloat[]{\includegraphics[scale=0.35]{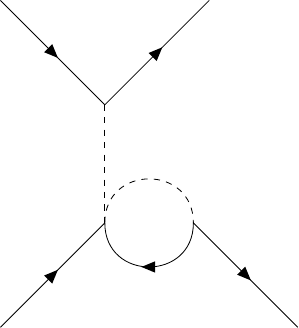}} \hspace*{8mm}
\subfloat[]{\includegraphics[scale=0.35]{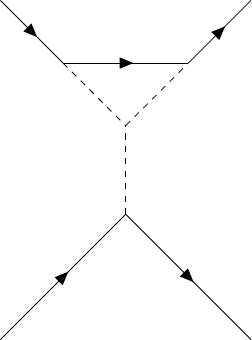}}\hspace*{8mm}
\subfloat[]{\includegraphics[scale=0.35]{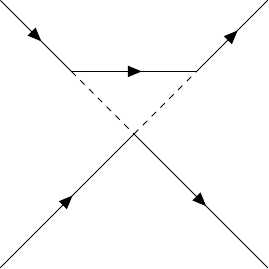}}
\caption{$t$ channel diagrams.}
\label{fig:tchanneldiagrams}
\end{figure}

For the analysis, we need to compute 21 Feynman diagrams. The diagrams that contribute to the $s$ channel are shown in figure \ref{fig:schanneldiagrams}, and those diagrams relevant for  the $t$ channel can be seen in figure \ref{fig:tchanneldiagrams}. The $u$ channel diagrams are analogous to the $t$ channel. The only difference corresponds to a permutation of isospin indexes of the external legs.

In previous works \cite{scattering1,scattering2,scattering3}, and because the sigma boson mass is much bigger than the pion mass, considering also the static limit approximation, the diagrams that contained sigma bosons were pinched, i.e., the sigma propagator was contracted to a point. The following limit was used
\begin{equation}
\dfrac{1}{p^2+m_\sigma^2} \longrightarrow \dfrac{1}{m_\sigma^2}.
\end{equation}
Pinching the sigma propagators simplify several diagrams, and all the calculations can be reduced to computing only five integrals. 

In the present analysis we present a detailed discussion of all diagrams, including full sigma propagators. This implies the calculation of several box diagrams which have not been considered in our previous articles.

Before going into the details of the calculation, it is useful to classify all integrals into different  types of integrals, which will be called $I_i$, where $i=1,\dots,9$ represents the $i-$type integral, except for the case of $I_{B1}$ and $I_{B2}$, that represents the integrals associated to the box diagrams 1 and 2, respectively. These integrals can be found in appendix \ref{ApendiceA}, where only terms up to the order  $\mathcal{O}(E^2)$ were considered. 

\subsection{\label{sec3b}Mathematical Methods }
For the purpose of computing the different loop corrections, essentially two methods were used. For some integrals the standard dimensional regularization was employed.


For the remaining integrals, which are the  majority of cases,  four dimensional hyper-spherical coordinates were used. In particular, the calculations of the first box diagram can be found in appendix \ref{ApendiceB}. Based on \cite{esfericas}, the set of coordinates used were
\begin{equation}
\begin{cases}
x_0= r\cos\t_1                  \\[7pt]
x_1= r\sin\t_1\cos\t_2          \\[7pt]
x_2= r\sin\fii\sin\t_1\sin\t_2  \\[7pt]
x_3= r\cos\fii\sin\t_1\sin\t_2,
\end{cases}
\end{equation}
where $0\leq \fii\leq 2\pi$, $0\leq \t_i\leq \pi$, $0\leq r< \infty$, and the jacobian of the transformation given by $J=r^3\sin^2\t_1\sin\t_2$. 

Along with the change of coordinates, it is important to note that a frame of reference can be set, without loss of generality, where the four momenta $p$ takes the form $p_\mu=\qty(m_\pi,\vb{0})$,  selecting a privileged direction according to
\begin{equation}
k\cdot p=m_\pi r\cos\t_1.
\end{equation}

\subsection{\label{sec3c}Isospin Projections}
Because of the associated Feynman rules, all the integrals emerging from the diagrams have a determined isospin structure. These structures are simplified when isospin projection operators acting on the different integrals are used. 


Using the projection operators \eqref{ProjOp0}, \eqref{ProjOp1}, and \eqref{ProjOp2},
all projections can be easily obtained. The numerical factors associated to each one of them can be seen in the following table \ref{tab:IsospinTable}.

\begin{table}[h]
\renewcommand{\arraystretch}{1.5}
\setlength{\tabcolsep}{3mm}
\centering
\begin{tabular}{c|c|c|c}
Isospin Structure & $P_0$ & $P_1$ & $P_2$\\\hline\hline
$7\d_{\a\g}\d_{\b\e}+2\d_{\a\b}\d_{\g\e}+2\d_{\a\e}\d_{\b\g}$ & $15$ & $-15$ & $45$ \\
$7\d_{\a\b}\d_{\g\e}+2\d_{\a\g}\d_{\b\e}+2\d_{\a\e}\d_{\b\g}$ & $25$ & $0$ & $20$\\
$\d_{\a\b}\d_{\g\e}+\d_{\a\g}\d_{\b\e}+\d_{\a\e}\d_{\b\g}$ & $5$ & $0$ & $10$\\
$\d_{\a\b}\d_{\g\e}$& $3$ & $0$ & $0$\\
$\d_{\a\g}\d_{\b\e}$& $1$ & $-3$ & $5$\\
$7\d_{\a\e}\d_{\b\g}+2\d_{\a\b}\d_{\g\e}+2\d_{\a\g}\d_{\b\e}$ & $15$ & $15$ & $45$\\
$\d_{\a\e}\d_{\b\g}$ & $1$ & $3$ & $5$\\
$\d_{\a\b}\d_{\g\e}+\d_{\a\e}\d_{\b\g}+\d_{\a\g}\d_{\b\e}$ & $5$ & $0$ & $10$\\
\end{tabular}
\caption{Table of factors obtained from the different isospin structure when acting with the isospin projectors. }
\label{tab:IsospinTable}
\end{table}

After this computation, it is interesting to note that both the $t$ and $u$ channel give the same results, except for an opposite sign. 
Therefore, our analysis leads to the following scattering amplitudes in the three isospin channels. 
\begin{align}
T^0&= 3A(s,t,u)+2 A(t,s,u),\label{CanalI0}\\
T^1&=0\label{CanalI1},\\
T^2&= 2A(t,s,u).\label{CanalI2}
\end{align}

Using expressions \ref{CanalI0}, \ref{CanalI1} and \ref{CanalI2}, along with the projections reported in table \ref{tab:IsospinTable}, it is found that the scattering amplitudes get the following form.
\begin{widetext}
\begin{align}
T^0&=32 \lambda ^8 v^4\qty(\dfrac{9}{2}I_{B1}+I_{B2}+I_9)+I_{3} \left(\frac{144 \lambda ^8 v^4}{4 m_{\pi}^2+m_{\sigma }^2}+120 \lambda ^6 v^2\right)+I_{4} \left(\frac{32 \lambda ^8 v^4}{m_{\sigma }^2}+80 \lambda ^6 v^2\right)+I_{6} \left(\frac{432 \lambda ^8 v^4}{4 m_{\pi}^2+m_{\sigma }^2}+72 \lambda ^6 v^2\right)\nonumber\\[5pt]
&\hspace*{1cm}+I_{5} \left(\frac{96 \lambda ^8 v^4}{m_{\sigma }^2}+16 \lambda ^6 v^2\right)+I_{7} \left(\frac{216 \lambda ^6 v^3}{\left(4 m_{\pi}^2+m_{\sigma }^2\right){}^2}+\frac{48 \lambda ^6 v^3}{M^4}\right)+I_{8} \left(\frac{72 \lambda ^6 v^2}{4 m_{\pi}^2+m_{\sigma }^2}+\frac{16 \lambda ^6 v^2}{m_{\sigma }^2}\right)\nonumber\\[5pt]
&\hspace*{1cm}+I_2 \left(300 \lambda ^4+\frac{432 \lambda ^8 v^4}{\left(4 m_{\pi}^2+m_{\sigma }^2\right){}^2}+\frac{360 \lambda ^6 v^3}{4 m_{\pi}^2+m_{\sigma }^2}\right)+I_1 \left(120 \lambda ^4+\frac{96 \lambda ^8 v^4}{m_{\sigma }^4}+\frac{80 \lambda ^6 v^3}{m_{\sigma }^2}\right),\label{T0Completo}\\[1.2cm]
T^2&= 160\lambda^8v^4\qty(I_{B2}+I_9)+I_{4} \left(\frac{160 \lambda ^8 v^4}{m_{\sigma }^2}+160 \lambda ^6 v^2\right)+I_{5} \left(\frac{480 \lambda ^8 v^4}{m_{\sigma }^2}+80 \lambda ^6 v^2\right)\nonumber\\[5pt]
&\hspace*{1cm}+\frac{240 \lambda ^6 v^3}{m_{\sigma }^4}I_{7}+\frac{80 \lambda ^6 v^2}{m_{\sigma }^2}I_{8}+I_1 \left(360 \lambda ^4+\frac{480 \lambda ^8 v^4}{m_{\sigma }^4}+\frac{400 \lambda ^6 v^3}{m_{\sigma }^2}\right) \label{T2Completo}.
\end{align}
\end{widetext}



 In order to obtain the scattering lengths $a_0^I$, using Eq.~\eqref{eq:a0I}, we get  
\begin{eqnarray}
a_0^0(E)	&=&a_0^0(\text{exp})+\frac1{32\pi}T^0,\nonumber\\
a_0^2(E)	&=&a_0^2(\text{exp})+\frac1{32\pi}T^2, \label{a0a2}
\end{eqnarray}
where $T^0$ and $T^2$ are given by Eq.~\eqref{T0Completo} and Eq.~\eqref{T2Completo}, respectively.  The experimental values at tree level are determined by \cite{Peyaud:2009zz} $a_0^0(\text{exp})=0.217$ and $a_0^2(\text{exp})=-0.041$.

\begin{figure}[H]
\centering
\includegraphics[scale=0.5]{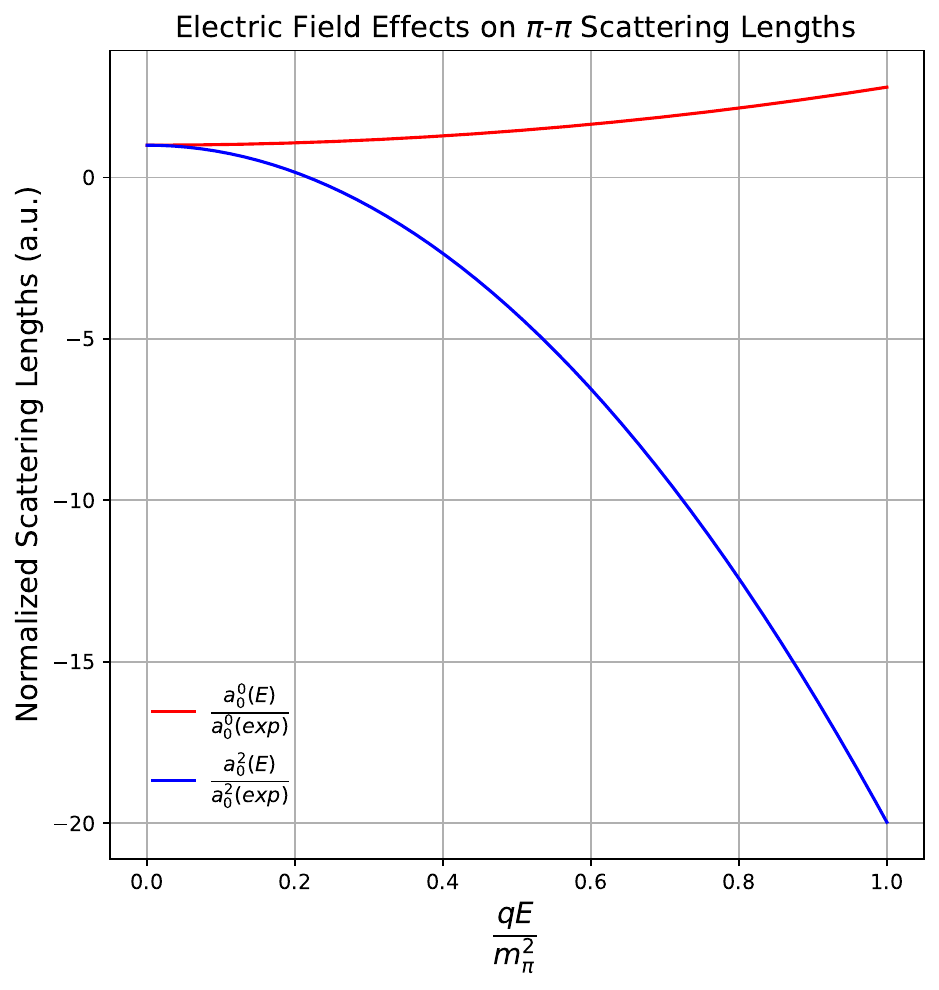}
\caption{Behaviour of the normalized scattering lengths as a function of $qE/m_\pi^2$. Red lines represent $a_0^0(E)/a_0^0(\text{exp})$, while blue lines represent $a_0^2(E)/a_0^2(\text{exp})$.}
\label{fig:Plot}
\end{figure}

In order to discuss the behavior of scattering lengths in the presence of an electric field, we will employ Eq.~\eqref{a0a2} normalizing to experimental values. We will use the following parameters $m_{\pi}=\SI{140}{\mega\electronvolt}$, $m_\sigma=\SI{550}{\mega\electronvolt}$, $v=\SI{89}{\mega\electronvolt}$, and $\lambda^2=4.26$, obtaining the following plot \ref{fig:Plot}.

\section{\label{sec4}Conclusions}

We have presented an analytic  calculation of $\pi$-$\pi$ scattering lengths within the linear sigma model at the one-loop level in the isospin channels $I=\{0,2\}$, as a function of the external electric field intensity. We provide a plot illustrating the calculation of scattering lengths $a_0^0$ and $a_0^2$ as a function of the electric field. It is evident that $a_0^0$ increases with the electric field, while $a_0^2$ decreases. This behavior is interesting, as it contrasts sharply with the effect of an external magnetic field. Such opposition between electric and magnetic fields is also observed in the calculation of renormalons \cite{propagadorelectrico1}. A significant novelty of this calculation is our inclusion of box diagrams, which have been fully computed, unlike in previous works \cite{scattering1,scattering2,scattering3}.

\begin{acknowledgments}
R. C. acknowledges support from ANID/CONICYT FONDECYT Regular (Chile) under Grant No. 1220035. M. L. acknowledges support from ANID/CONICYT FONDECYT Regular (Chile) under Grant No. 1241436 and 1220035. R. Z. acknowledges support from ANID/CONICYT FONDECYT Regular (Chile) under Grant No. 1241436 and 1220035.
\end{acknowledgments}

\appendix
\section{\label{ApendiceA}Types of Integrals}
As stated in section \ref{sec3a}, by analyzing all the integrals found from the Feynman diagrams, we can classify them into several types, as shown below. 
\begin{enumerate}[label={}, leftmargin=*]
\item $\ds I_1 \equiv \int\dfrac{\dd[4]{k}}{(2\pi)^4}D(k)^2$,
\item $\ds I_2 \equiv\int\dfrac{\dd[4]{k}}{(2\pi)^4}D(k)D(k-2p)$,
\item $\ds I_3 \equiv\int\dfrac{\dd[4]{k}}{(2\pi)^4}S(k)D(p+k)D(p-k)$,
\item $\ds I_4 \equiv\int\dfrac{\dd[4]{k}}{(2\pi)^4}S(k)D(p-k)^2$,
\item $\ds I_5 \equiv\int\dfrac{\dd[4]{k}}{(2\pi)^4}S(p-k)^2D(k)$,
\item $\ds I_6 \equiv\int\dfrac{\dd[4]{k}}{(2\pi)^4}S(p+k)S(p-k)D(k)$,
\item $\ds I_7 \equiv\int\dfrac{\dd[4]{k}}{(2\pi)^4}D(k)$,
\item $\ds I_8 \equiv\int\dfrac{\dd[4]{k}}{(2\pi)^4}S(p\mp k)D(k)$,
\item $\ds I_9 \equiv\int\dfrac{\dd[4]{k}}{(2\pi)^4}S(p-k)^2D(k)^2$,
\item $\ds I_{B1} \equiv\int\dfrac{\dd[4]{k}}{(2\pi)^4}S(p+k)S(p-k)D(k)^2$,
\item $\ds I_{B2} \equiv\int\dfrac{\dd[4]{k}}{(2\pi)^4}S(p-k)^2D(k)D(2p-k)$,
\end{enumerate}
with $D(p)$ and $S(p)$ is defined in Eq.~\eqref{PiPropagator} and Eq.~\eqref{libre} respectively.

Also, in $I_8$, the change of sign does not produce a new class of integrals.

\section{\label{ApendiceB}Box Diagram Calculation}
As mentioned in section \ref{sec3b}, one of the innovations of this work is the calculation of the full box diagrams. Let's begin by analyzing the first one of them. In the $s$ channel, we have 
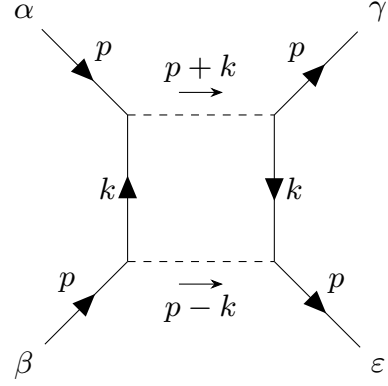
\begin{figure}[H]
\centering
\begin{tikzpicture}[scale=1.3, every node/.style={scale=1.3}]
\begin{feynman}
\vertex(a);
\vertex[right of=a](b);
\vertex[below of=b](c);
\vertex[below of=a](d);
\vertex[above left of=a](a1){$\a$};
\vertex[above right of=b](b1){$\g$};
\vertex[below right of=c](c1){$\e$};
\vertex[below left of=d](d1){$\b$};
\diagram{
	(a1)--[fermion, edge label=$p$](a);
	(a)--[scalar, momentum={[arrow shorten=0.35]$p+k$}](b);
	(b)--[fermion, edge label=$p$](b1);
	(d)--[fermion, edge label=$k$](a);
	(b)--[fermion, edge label=$k$](c);
	(d1)--[fermion, edge label=$p$](d);
	(d)--[scalar, momentum'={[arrow shorten=0.35]$p-k$}](c);
	(c)--[fermion, edge label=$p$](c1);
};
\end{feynman}
\end{tikzpicture}
\caption{Box diagram 1 associated to the $s$ channel. }
\end{figure}

By means of an adequate isospin parametrization, together with the Feynman rules, the associated integral $I_{B1}$ gets the following form
\begin{equation}
I_{B1}= 16\lambda^8v^4\d_{\a\b}\d_{\g\e}\int\dfrac{\dd[4]{k}}{(2\pi)^4}\: D(k)^2S(k+p)S(k-p).
\end{equation}

After expanding the propagators up to $\mathcal{O}(E^2)$, we get 
\begin{widetext}
\begin{align}
&D(k)^2S(p+k)S(p-k)= \dfrac{1}{\qty(k^2+m_{\pi}^2)^2\qty(\qty(p+k)^2+m_\sigma^2)\qty(\qty(p-k)^2+m_\sigma^2)}\nonumber\\
&\hspace*{0.5cm}+2q^2E^2\Bigg[\dfrac{2\qty(k_\perp^2+m_{\pi}^2)}{\qty(k^2+m_{\pi}^2)^5\qty(\qty(p+k)^2+m_\sigma^2)\qty(\qty(p-k)^2+m_\sigma^2)}-\dfrac{1}{\qty(k^2+m_{\pi}^2)^4\qty(\qty(p+k)^2+m_\sigma^2)\qty(\qty(p-k)^2+m_\sigma^2)}\Bigg]+O(E^3) .\label{B1PropagatorExpansion}
\end{align}
\end{widetext}

To continue our calculation, from Eq.~\eqref{B1PropagatorExpansion},  we have to deal with three different integrals. These calculations were performed using the hyper-spherical coordinates introduced in section \ref{sec3b}. We obtain
\begin{widetext}
\begin{align}
\mathcal{I}_1&= \int \dfrac{1}{\qty(r^2+m_{\pi}^2)^2\qty(r^2+2m_{\pi}r\cos\t_1+m_{\pi}^2+m_\sigma^2)\qty(r^2-2m_{\pi}r\cos\t_1+m_{\pi}^2+m_\sigma^2)}\cdot r^3\dd{r}\dd{\Omega_4}, \\[5mm]
\mathcal{I}_2&= \int \dfrac{\qty(r\sin\fii\sin\t_1\sin\t_2)^2+\qty(r\sin\t_1\cos\t_2)^2+m^2}{\qty(r^2+m_{\pi}^2)^5\qty(r^2+2m_{\pi}r\cos\t_1+m_{\pi}^2+m_\sigma^2)\qty(r^2-2m_{\pi}r\cos\t_1+m_{\pi}^2+m_\sigma^2)} \cdot r^3\dd{r}\dd{\Omega_4}, \\[5mm]
\mathcal{I}_3&= \int \dfrac{1}{\qty(r^2+m_{\pi}^2)^4\qty(r^2+2m_{\pi}r\cos\t_1+m_{\pi}^2+m_\sigma^2)\qty(r^2-2m_{\pi}r\cos\t_1+m_{\pi}^2+m_\sigma^2)} \cdot r^3\dd{r}\dd{\Omega_4},
\end{align}
\end{widetext}
where $\dd{\Omega_4}=\sin^2\t_1\sin\t_2\dd{\t_1}\dd{\t_2}$ represents the angular measure. It is important to note that the angular integration can be carried out without major problems. After performing all the integrals, we find the following results.
\begin{widetext}
\begin{align}
\mathcal{I}_1&= \dfrac{2 m_{\pi} \left(2 m_{\pi}^2+m_{\sigma }^2\right) \left(\tanh ^{-1}\frac{m_{\sigma }^2}{\sqrt{4 m_{\pi}^4+m_{\sigma }^4}}+\ln\frac{\sqrt{4 m_{\pi}^4+m_{\sigma }^4}+2 m_{\pi}^2-m_{\sigma }^2}{\sqrt{4 m_{\pi}^4+m_{\sigma }^4}-2 m_{\pi}^2+m_{\sigma }^2}\right)-\sqrt{4 m_{\pi}^6+4 m_{\pi}^4 m_{\sigma }^2+m_{\pi}^2 m_{\sigma }^4+m_{\sigma }^6} \ln \left(\frac{\sqrt{m_{\pi}^2+m_{\sigma }^2}+m}{\sqrt{m_{\pi}^2+m_{\sigma }^2}-m_{\pi}}\right)}{16 \pi ^2 m_{\pi} m_{\sigma }^4 \sqrt{4 m_{\pi}^4+m_{\sigma }^4}},\label{I1Box1}\\[5mm]
\mathcal{I}_2&= \dfrac{1}{2304 \pi ^2 m_{\pi }^{12} m_{\sigma }^{10} \left(m_{\sigma }^4+4 m_{\pi }^4\right){}^{7/2}}\Bigg[
m_{\sigma }^{10} \left(-8 m_{\pi }^2 m_{\sigma }^2-3 m_{\sigma }^4+18 m_{\pi }^4\right) \left(m_{\sigma }^4+4 m_{\pi }^4\right){}^{7/2}\nonumber\\
&\hspace*{0.2cm}+m_{\sigma }^4\sqrt{m_{\sigma }^4+4 m_{\pi }^4}  \bigg(-1536 m_{\pi }^{20} m_{\sigma }^2+480 m_{\pi }^{18} m_{\sigma }^4-1536 m_{\pi }^{16} m_{\sigma }^6-160 m_{\pi }^{14} m_{\sigma }^8-420 m_{\pi }^{12} m_{\sigma }^{10}\nonumber\\
&\hspace*{5cm}+288 m_{\pi }^{10} m_{\sigma }^{12}-54 m_{\pi }^8 m_{\sigma }^{14}+96 m_{\pi }^6 m_{\sigma }^{16}+18 m_{\pi }^4 m_{\sigma }^{18}+8 m_{\pi }^2 m_{\sigma }^{20}+3 m_{\sigma }^{22}+768 m_{\pi }^{22}\bigg)\nonumber\\
&\hspace*{0.2cm}+48 m_{\pi }^{11} \sqrt{m_{\sigma }^2+m_{\pi }^2} \left(m_{\pi }^2-2 m_{\sigma }^2\right) \left(m_{\sigma }^4+4 m_{\pi }^4\right){}^{7/2} \ln \left(\frac{2 m_{\pi } \left(\sqrt{m_{\sigma }^2+m_{\pi }^2}+m_{\pi }\right)}{m_{\sigma }^2}+1\right)\nonumber\\
&\hspace*{0.2cm}-24 m_{\pi }^{12} \left(-384 m_{\pi }^{14} m_{\sigma }^2-64 m_{\pi }^{12} m_{\sigma }^4-256 m_{\pi }^{10} m_{\sigma }^6-224 m_{\pi }^8 m_{\sigma }^8-8 m_{\pi }^6 m_{\sigma }^{10}-114 m_{\pi }^4 m_{\sigma }^{12}+8 m_{\pi }^2 m_{\sigma }^{14}-9 m_{\sigma }^{16}+256 m_{\pi }^{16}\right)\nonumber\\
&\hspace*{7cm}\cdot \left(2 \tanh ^{-1}\frac{m_{\sigma }^2}{\sqrt{m_{\sigma }^4+4 m_{\pi }^4}}-\ln \frac{m_{\sigma }^2+\sqrt{m_{\sigma }^4+4 m_{\pi }^4}-2 m_{\pi }^2}{-m_{\sigma }^2+\sqrt{m_{\sigma }^4+4 m_{\pi }^4}+2 m_{\pi }^2}\right)
\Bigg],\label{I2Box1} \\[5mm]
\mathcal{I}_3&= \dfrac{1}{96 \pi ^2 m_{\pi }^4 m_{\sigma }^8 \left(m_{\sigma }^4+4 m_{\pi }^4\right){}^3}\Bigg[
m_{\sigma }^4 \left(-32 m_{\pi }^8 m_{\sigma }^4-24 m_{\pi }^6 m_{\sigma }^6+2 m_{\pi }^4 m_{\sigma }^8-6 m_{\pi }^2 m_{\sigma }^{10}+m_{\sigma }^{12}-96 m_{\pi }^{12}\right)\nonumber\\
&\hspace*{0.2cm}-6 m_{\pi }^3 \sqrt{m_{\sigma }^2+m_{\pi }^2} \left(m_{\sigma }^4+4 m_{\pi }^4\right){}^3 \ln \left(\frac{2 m_{\pi } \left(\sqrt{m_{\sigma }^2+m_{\pi }^2}+m_{\pi }\right)}{m_{\sigma }^2}+1\right)\nonumber\\
&\hspace*{0.2cm}+12 m_{\pi }^4 \sqrt{m_{\sigma }^4+4 m_{\pi }^4} \left(16 m_{\pi }^8 m_{\sigma }^2+16 m_{\pi }^6 m_{\sigma }^4+12 m_{\pi }^4 m_{\sigma }^6+3 m_{\sigma }^{10}+32 m_{\pi }^{10}\right) \left(\tanh ^{-1}\frac{m_{\sigma }^2}{\sqrt{m_{\sigma }^4+4 m_{\pi }^4}}+\tanh ^{-1}\frac{2 m_{\pi }^2-m_{\sigma }^2}{\sqrt{m_{\sigma }^4+4 m_{\pi }^4}}\right)
\Bigg].\label{I3Box1}
\end{align}
\end{widetext}

With expressions \eqref{I1Box1}, \eqref{I2Box1} and \eqref{I3Box1}, these can be combined and simplified to obtain an analytical result for the total loop. In fact, from the expansion of the propagators, it is found that we can write
\begin{equation}
I_{B1}=-16\lambda^8v^4\d_{\a\b}\d_{\g\e}\qty[\mathcal{I}_1+2q^2E^2\qty(2\mathcal{I}_2-\mathcal{I}_3)]
\label{IB1Total}
\end{equation}
Therefore, and because we are interested in the electric corrections, the vacuum term is not considered. From this, after replacing \eqref{I2Box1} and \eqref{I3Box1} into \eqref{IB1Total}, the terms proportional to $\qty(qE)^2$ are given by the following expression. 
\begin{widetext}
\begin{align}
I_{B1}&= -\dfrac{\lambda^8v^4\d_{\a\b}\d_{\g\e}\qty(qE)^2}{36 \pi ^2 m_{\pi }^{12} m_{\sigma }^{10} \left(m_{\sigma }^4+4 m_{\pi }^4\right){}^{7/2}}\Bigg[
12 m_{\pi }^8 m_{\sigma }^6 \left(2 m_{\pi }^4 m_{\sigma }^4+6 m_{\pi }^2 m_{\sigma }^6-m_{\sigma }^8+24 m_{\pi }^8\right) \left(m_{\sigma }^4+4 m_{\pi }^4\right){}^{3/2}\nonumber\\
&\hspace*{0.2cm}+m_{\sigma }^{10} \left(-8 m_{\pi }^2 m_{\sigma }^2-3 m_{\sigma }^4+18 m_{\pi }^4\right) \left(m_{\sigma }^4+4 m_{\pi }^4\right){}^{7/2}\nonumber\\
&\hspace*{0.2cm}+\sqrt{m_{\sigma }^4+4 m_{\pi }^4} m_{\sigma }^4 \bigg(-1536 m_{\pi }^{20} m_{\sigma }^2+480 m_{\pi }^{18} m_{\sigma }^4-1536 m_{\pi }^{16} m_{\sigma }^6-160 m_{\pi }^{14} m_{\sigma }^8-420 m_{\pi }^{12} m_{\sigma }^{10}+288 m_{\pi }^{10} m_{\sigma }^{12}\nonumber\\
&\hspace*{8cm}-54 m_{\pi }^8 m_{\sigma }^{14}+96 m_{\pi }^6 m_{\sigma }^{16}+18 m_{\pi }^4 m_{\sigma }^{18}+8 m_{\pi }^2 m_{\sigma }^{20}+3 m_{\sigma }^{22}+768 m_{\pi }^{22}\bigg)\nonumber\\
&\hspace*{0.2cm}+24 m_{\pi }^{11} \sqrt{m_{\sigma }^2+m_{\pi }^2} \left(2 m_{\pi }^2-m_{\sigma }^2\right) \left(m_{\sigma }^4+4 m_{\pi }^4\right){}^{7/2} \ln \left(\frac{2 m_{\pi } \left(\sqrt{m_{\sigma }^2+m_{\pi }^2}+m_{\pi }\right)}{m_{\sigma }^2}+1\right)\nonumber\\
&\hspace*{0.2cm}-144 m_{\pi }^{12} m_{\sigma }^2 \left(16 m_{\pi }^8 m_{\sigma }^2+16 m_{\pi }^6 m_{\sigma }^4+12 m_{\pi }^4 m_{\sigma }^6+3 m_{\sigma }^{10}+32 m_{\pi }^{10}\right) \left(m_{\sigma }^4+4 m_{\pi }^4\right)\nonumber\\
&\hspace*{6cm}\cdot \left(\tanh ^{-1}\frac{m_{\sigma }^2}{\sqrt{m_{\sigma }^4+4 m_{\pi }^4}}+\tanh ^{-1}\frac{2 m_{\pi }^2-m_{\sigma }^2}{\sqrt{m_{\sigma }^4+4 m_{\pi }^4}}\right)\nonumber\\
&\hspace*{0.2cm}+24 m_{\pi }^{12} \left(-384 m_{\pi }^{14} m_{\sigma }^2-64 m_{\pi }^{12} m_{\sigma }^4-256 m_{\pi }^{10} m_{\sigma }^6-224 m_{\pi }^8 m_{\sigma }^8-8 m_{\pi }^6 m_{\sigma }^{10}-114 m_{\pi }^4 m_{\sigma }^{12}+8 m_{\pi }^2 m_{\sigma }^{14}-9 m_{\sigma }^{16}+256 m_{\pi }^{16}\right)\nonumber\\
&\hspace*{6cm}\cdot \left(\ln \frac{m_{\sigma }^2+\sqrt{m_{\sigma }^4+4 m_{\pi }^4}-2 m_{\pi }^2}{-m_{\sigma }^2+\sqrt{m_{\sigma }^4+4 m_{\pi }^4}+2 m_{\pi }^2}-2 \tanh ^{-1}\frac{m_{\sigma }^2}{\sqrt{m_{\sigma }^4+4 m_{\pi }^4}}\right)
\Bigg]
\end{align}
\end{widetext}

With this correction, the second box diagram can be computed in a completely analogous way. However, the calculation for $I_{B2}$ is much longer, because in that case, five integrals must be solved to get the needed analytical result.

\bibliography{ourbibliography}
\end{document}